\documentclass[3p,times]{elsarticle}

\usepackage{ecrc}
\volume{00}
\firstpage{1}

\journalname{Nuclear Physics A}

\runauth{E. Iancu}
\jid{nupha}
\usepackage{latexsym,bm,amsmath,amssymb,amsfonts}

\begin{document}

\begin{frontmatter}

\dochead{}

\title{Parton branching and medium--induced
radiation in a strongly coupled plasma}


\author{Edmond Iancu}

\address{CERN, Theory Division, CH-1211 Geneva, Switzerland}

\begin{abstract}
I review the parton picture at strong coupling emerging from the
gauge/gravity duality together with its consequences for the energy loss
and momentum broadening of a heavy quark moving through a strongly
coupled plasma.
\end{abstract}




\end{frontmatter}

\section{Motivation: Jet quenching at RHIC}


Some of the experimental discoveries at RHIC, notably the unexpectedly
large medium effects known as elliptic flow and jet quenching, led to the
suggestion that the deconfined QCD matter produced in the intermediate
stages of a heavy ion collision is strongly interacting. The phenomenon
of jet quenching is particularly intriguing in that sense, as this probes
the hadronic matter on a relatively hard scale (a few GeV), where QCD is
{\em a priori} expected to be weakly coupled, by asymptotic freedom. One
manifestation of this phenomenon is the `away jet suppression' in Au+Au
collisions: unlike in p+p or d+Au collisions, where the hard particles
typically emerge from the collision region as pairs of back--to--back
jets, in the Au+Au collisions one sees `mono--jet' events in which the
second jet is missing (at RHIC, `jet' means leading particle). This has
the following natural interpretation: the hard scattering producing the
jets has occurred near the edge of the interaction region, so that one of
the jets has escaped and triggered a detector, while the other one has
been deflected, slowed down, or even absorbed, via interactions in the
surrounding medium.

\begin{figure*}{ \centerline{
\includegraphics[width=.4\textwidth]{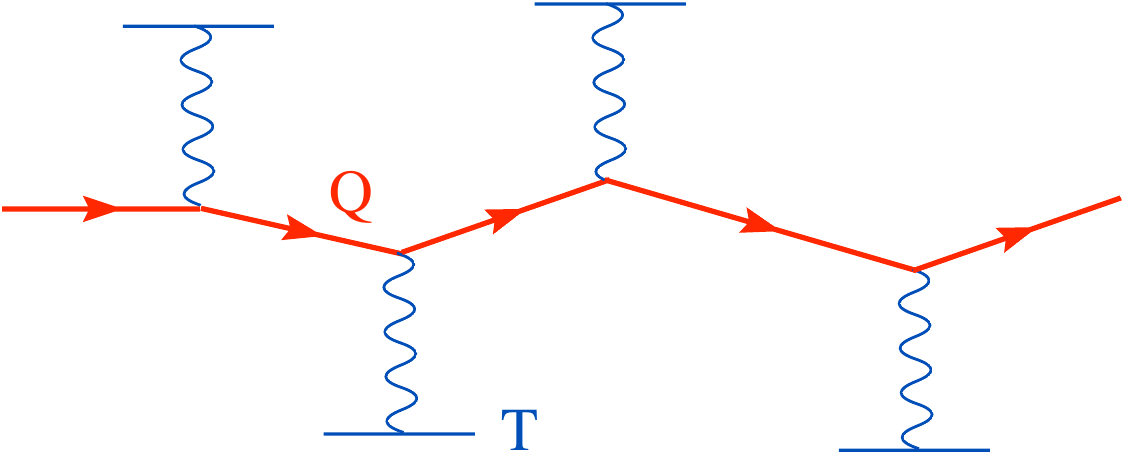}\qquad
\includegraphics[width=.4\textwidth]{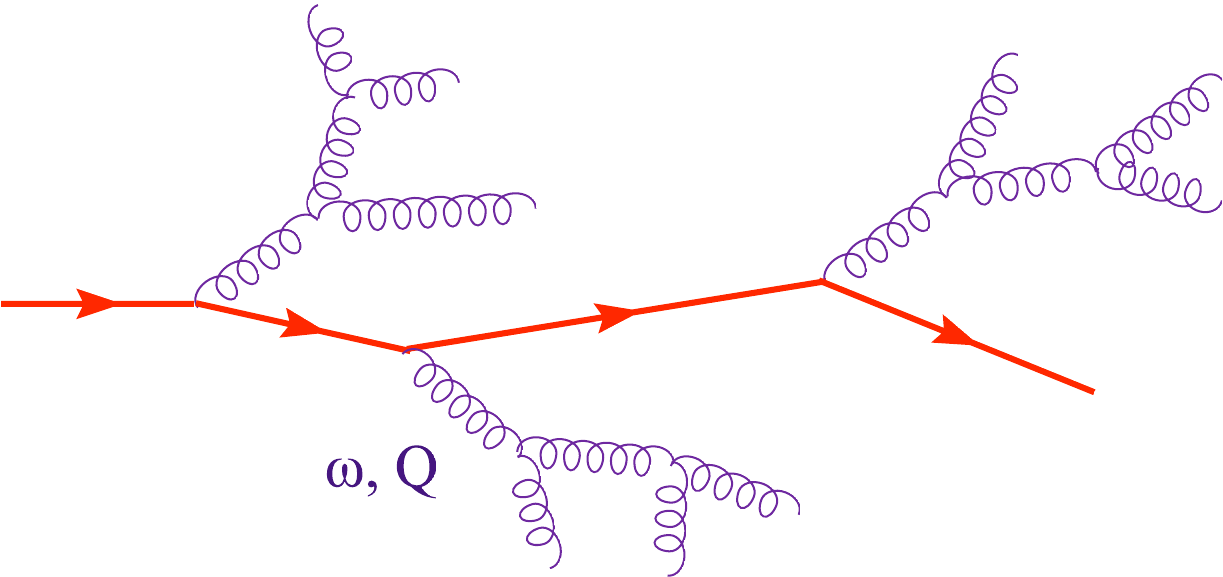}}
\caption{\sl\small Transverse momentum broadening for a heavy quark which
propagates through a quark--gluon plasma. Left: weak coupling (successive
scattering off medium constituents). Right: strong coupling (medium
induced parton branching).} \label{fig:quench}}
\end{figure*}

If this medium is composed of weakly interacting quasiparticles (quarks
and gluons), then the deflection of the hard jet is due to its successive
scattering off these quasiparticles, as illustrated in
Fig.~\ref{fig:quench} left. This leads to the following estimate for the
rate of transverse momentum broadening \cite{Wiedemann:2009sh}
 \begin{eqnarray}\label{qhat}
 \hat q\,\equiv\, \frac{{\rm d} \langle k_\perp^2\rangle}{{\rm d} t}
  \,\simeq\,\alpha_s N_c \,x G(x,Q^2),
 \end{eqnarray}
where $x G(x,Q^2)$ is the gluon distribution in the medium on the
resolution scale $Q^2\sim \langle k_\perp^2\rangle$ of the hard jet, as
produced via quantum evolution from the medium intrinsic scale up to $Q$.
For instance, if the medium is a finite--temperature plasma with
temperature $T$, then $x G\simeq n_q(T)\,x G_q +n_g(T)\,x G_g$, where
$n_{q,g}(T)\propto T^3$ are the quark and gluon densities in thermal
equilibrium and $x G_{q,g}(x,Q^2)$ are the gluon distributions produced
by a single quark, respectively gluon, on the scale $Q\gg T$. Some
typical values at RHIC are $T\!\sim \!0.4$~GeV and $Q\!\sim\! 2\div
12$~GeV. At the LHC, they become $T\!\sim\! 0.7$~GeV and $Q\sim 100$ GeV
(for actual jets). Assuming weak coupling, one can compute
Eq.~\eqref{qhat} in perturbation theory. But by doing that, one finds an
estimate $\hat q_{\rm pQCD}\simeq 0.5\div 1\, {\rm GeV}^2/{\rm fm}$ which
is considerably smaller (by almost one order of magnitude) then the value
extracted from the RHIC data ! This puzzle is comforted by the first data
for Pb+Pb collisions at LHC, which confirm the strong jet quenching
observed at RHIC and suggest that the medium effects can remain important
even for jets as hard as $Q\sim\! 100$~GeV.

A possible solution to this puzzle is that the deconfined QCD matter is
(relatively) strongly coupled. This would enhance the quantum evolution
from $T$ to $Q$ and also the interactions between the hard probe and the
medium. Note that there is not necessarily a conflict with asymptotic
freedom: to get an enhanced gluon distribution on the hard scale $Q$, it
is enough to have a stronger coupling at the lower end of the evolution,
that is, at the softer scale $T$. We have indeed $g(T)\sim 2$ for the
temperatures $T$ of interest at RHIC and LHC. Actually, it should be
possible to study some aspects of this evolution in lattice QCD at
finite--$T$ and thus verify the hypothesis of strong coupling
\cite{Latt}.

\section{Parton picture and jet quenching at strong coupling}

The previous discussion invites us to a better understanding of the
properties of the quark--gluon plasma (QGP) at relatively strong
coupling, that is, for $\alpha_s\!\equiv\! g^2/4\pi\! \sim \!1$. However,
even without the complications of confinement, the QCD calculations at
strong coupling remain notoriously difficult. (In particular, lattice QCD
cannot be used for real--time processes like transport phenomena and
scattering.) So it has become common practice to look to the ${\mathcal
N}=4$ supersymmetric Yang--Mills (SYM) theory, whose strong coupling
regime can be addressed within the AdS/CFT correspondence, for guidance
as to general properties of strongly coupled plasmas.

${\mathcal N}=4$ SYM has the `color' gauge symmetry SU$(N_c)$, so like
QCD, but differs from the latter in some other aspects: it has (maximal)
supersymmetry and hence conformal symmetry (the coupling $g$ is fixed)
and all the fields in its Lagrangian (gluons, scalars, and fermions)
transform in the adjoint representation of SU$(N_c)$. But these
differences are believed not to be essential for a study of the QGP phase
of QCD in the temperature range of interest for heavy ion collisions at
RHIC and LHC, that is, $2T_c\lesssim T \lesssim 5T_c$ with $T_c\simeq
170$~MeV the critical temperature for deconfinement. Indeed, lattice
studies indicate that the QGP itself is nearly conformal in this window.

\begin{figure*}[htb]
\centerline{\includegraphics[width=0.3\textwidth]{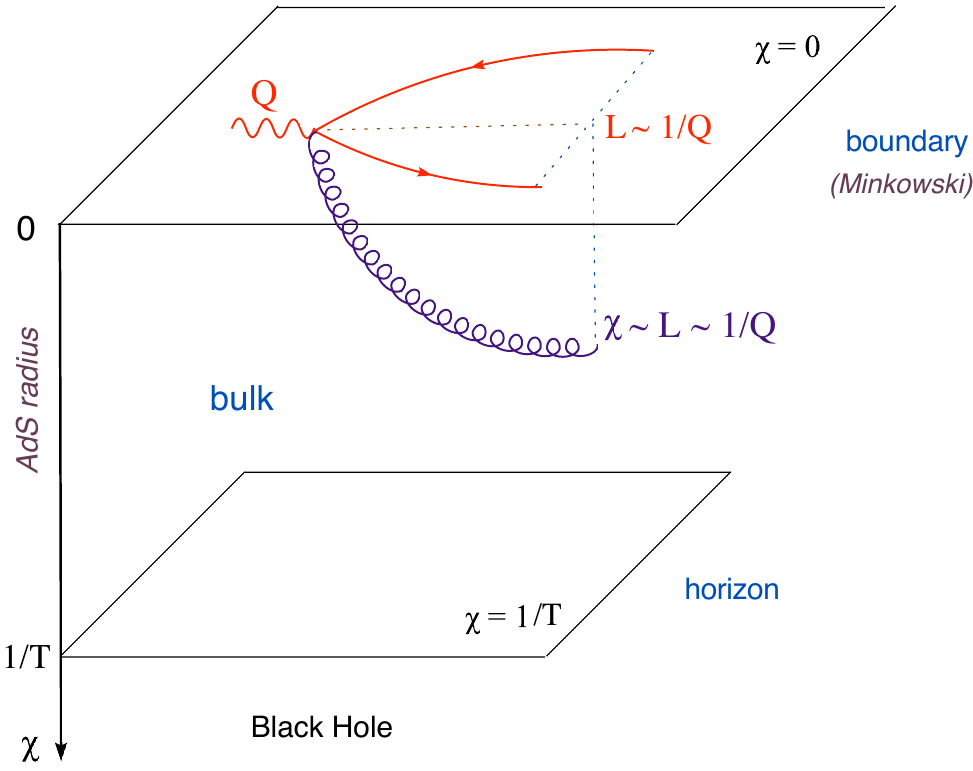} \qquad
\includegraphics[width=0.3\textwidth]{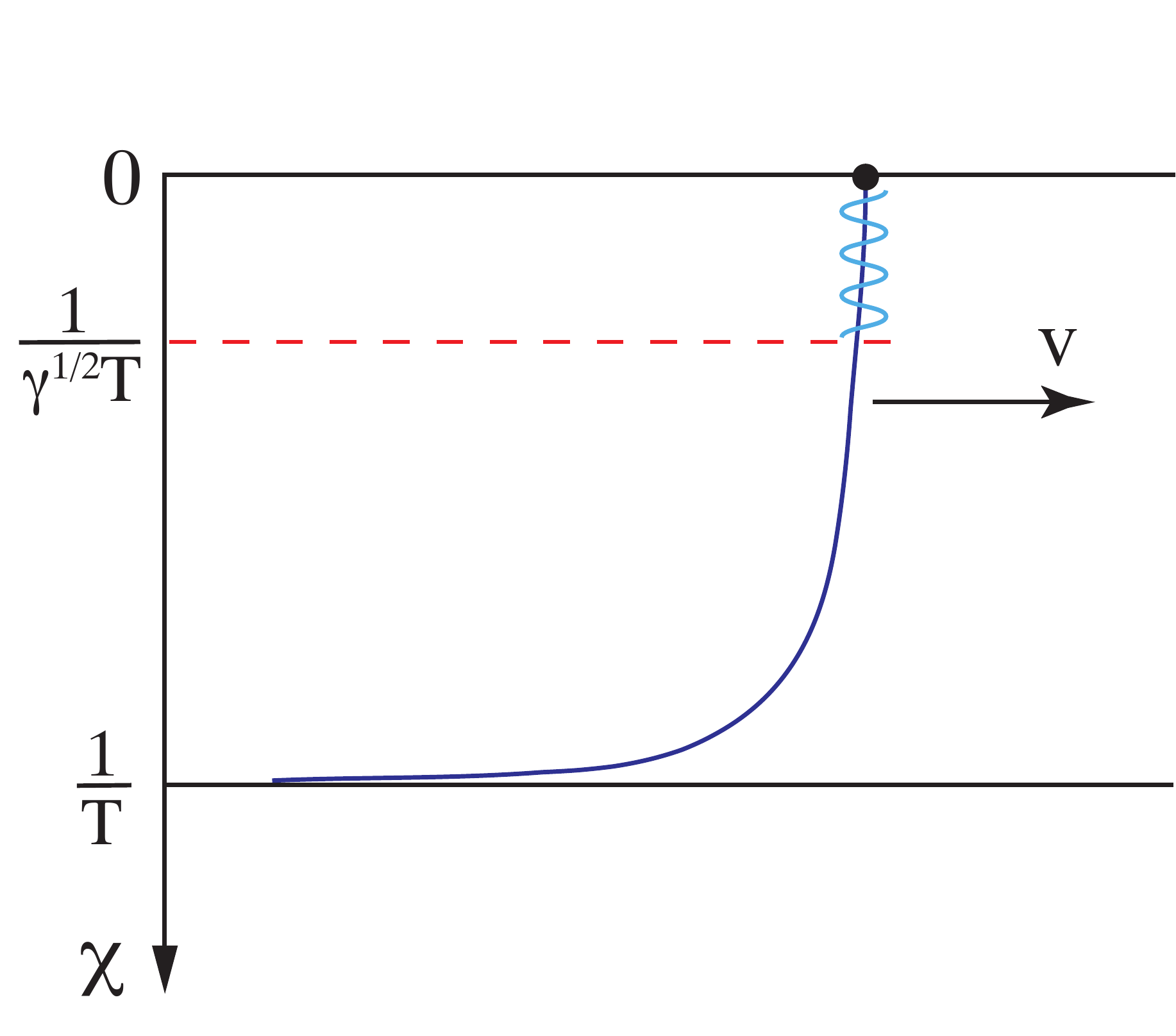}}
\caption{\small\sl  Left: dual description of a space--like current (the
figure shows the associated bulk excitation and its `shadow' on the
boundary). Right: the trailing string dual to a heavy quark in the plasma
(the dotted line marks the world--sheet horizon). \label{fig:WAVE}}
\end{figure*}

The AdS/CFT correspondence is the statement that the gauge theory
${\mathcal N}=4$ SYM is `dual' ({\em i.e.}, equivalent) to a special type
of string theory living in the $D=9+1$ space--time with AdS$_5\times S^5$
geometry. This duality is particularly useful in that it relates the
strong `t Hooft coupling limit $\lambda\equiv g^2N_c\to \infty$ (with
fixed $g\ll 1$) of the gauge theory to the weak--coupling limit of the
string theory, in which the latter reduces to classical gravity
(`supergravity' or SUGRA). The Anti-de-Sitter space--time AdS$_5$ can be
viewed as the product between the physical space--time with $D=3+1$ (the
Minkowski boundary of AdS$_5$) and a fifth (or `radial') dimension
$\chi$, with $0\le \chi<\infty$. Adding temperature on the gauge theory
side corresponds to embedding a black hole (BH) in AdS$_5$, with the BH
horizon located at a distance $\chi=1/T$ away from the Minkowski boundary
at $\chi=0$. (See Fig.~\ref{fig:WAVE} for some illustrations.)

Within this context, the gauge interactions between a hard probe and the
strongly coupled plasma are described as the gravitational interactions
between the dual bulk object (a string or a field) representing the hard
probe and the BH. This amounts to solving the classical equation of
motion for the propagation of the dual object in the AdS$_5$ BH
space--time. For instance, one can unveil the parton structure of the
plasma by studying `deep inelastic scattering' (DIS), that is, the
propagation of a virtual photon with space--like 4--momentum (see
Fig.~\ref{fig:WAVE}.a). The dual object in the bulk is a vector field
field $A_m$ (with $m=\mu$ or $\chi$) which obeys the Maxwell equations in
curved space--time:
 \begin{eqnarray}\label{maxwell} \partial_m\big(\sqrt{-g}g^{mp}g^{nq}
 F_{pq})\,=\,0\,,
 \qquad\mbox{where}\quad F_{mn}=\partial_m A_n-\partial_n A_m\,.
 \end{eqnarray}
The gravitational interactions are encoded in the 5D metric tensor
$g^{mn}$, which involves the horizon of the BH.

Similarly, in order to study the phenomenon of `jet quenching' for a
heavy quark, one needs the corresponding `dual' object in the bulk. This
is a string hanging down in AdS$_5$ with one endpoint attached to the
heavy quark on the boundary (see Fig.~\ref{fig:WAVE}.b). The average
string profile can be obtained by solving the corresponding equation of
motion (the Nambu--Goto equation) in the background of the BH. Then the
rate ${\rm d}E/{\rm d} t$ for energy loss is computed as the flux of
energy down the string. Also, the momentum broadening follows by studying
the (small) fluctuations of the string along this average profile. These
fluctuations reflect stochastic phenomena in the physical problem, whose
precise nature will be discussed later on. This string problem is
mathematically involved in the general case, but it has been fully
carried out in one special situation, known as the `trailing string'
\cite{Energyloss,Broadening}: this is the steady situation where the
quark (and hence the string as a whole) moves at constant velocity under
the action of an external force, which supplies the energy that is
continuously lost towards the plasma/BH.

It turns out that the two problems alluded to above --- DIS off the
plasma and the quenching of a heavy quark --- are connected at a deeper
physical level: the partonic picture of the strongly--coupled plasma, as
emerging from DIS \cite{HIM2}, enables us to propose a physical
interpretation for the AdS/CFT results for the trailing string, and also
a rapid derivation of these results to parametric accuracy \cite{String}.
The key observation is that both phenomena are controlled by the same
fundamental scale : a (space--like) virtuality scale that we shall refer
to as the `saturation momentum' $Q_s$. This scale is an intrinsic
property of the medium (more precisely, of its parton distribution),
although it also depends upon the energy of the `projectile' (the hard
probe) --- the latter only fixes the energy resolution at which one
probes the partonic distribution in the `target' (the plasma).

Specifically, consider a space--like photon with 4--momentum
$q^\mu=(\omega,0,0,q)$ and virtuality $Q^2\equiv q^2-\omega^2 \gg T^2$
which propagates through the plasma. By solving the AdS equation
\eqref{maxwell}, one finds two physical regimes: \texttt{(i)} a {\em low
energy} regime at $\omega \ll Q^3/T^2$, where there is essentially no
interaction between the bulk field $A_m$ and the BH (physically, this
means that the plasma looks transparent to the virtual photon), and
\texttt{(ii)} a {\em high energy} regime at $\omega \gtrsim Q^3/T^2$,
where the field feels the attraction of the BH and eventually falls into
it ({\em i.e.}, the photon is completely absorbed by the plasma). This
strong energy dependence of the results is natural in the context of
gravity, since gravitational interactions are proportional to the energy.
The physical interpretation of these results back in the original gauge
theory is more subtle and relies on the `UV/IR correspondence'. This
correspondence states that the radial penetration $\chi$ of the bulk
field in $AdS_5$ is proportional to the transverse size $L$ of the
partonic fluctuation of the virtual photon on the boundary (see
Fig.~\ref{fig:WAVE}.a). Consider e.g. a space--like photon in the vacuum:
energy--momentum conservation implies that this photon cannot decay into
on--shell quanta, but only fluctuate into a virtual partonic fluctuation
with lifetime $\Delta t\sim {\omega}/{Q^2}$ and transverse size $L\sim
1/Q$, as determined by the uncertainty principle. And indeed, in the low
energy regime above mentioned, the solution $A_m$ to Eq.~\eqref{maxwell}
is found to penetrate in AdS$_5$ (via diffusion:
$\chi\sim\sqrt{t/\omega}$\,) up to a maximal distance $\chi\sim 1/Q$,
which is reached after a time ${\omega}/{Q^2}$.

To similarly understand the high energy regime and the critical value
$\omega \sim Q^3/T^2$ separating between the two regimes, one needs an
additional piece of information: when viewing the equation of motion for
the bulk field $A_m$ through the prism of the UV/IR correspondence, one
concludes that the strongly--coupled plasma acts on partonic fluctuations
with a {\em tidal force} $F\sim T^2$ which pulls the partons apart. The
partons become nearly on--shell when the mechanical work $W= F\,\Delta t$
provided by this force during the lifetime $\Delta t\sim {\omega}/{Q^2}$
of the fluctuation is large enough to compensate the energy deficit $\sim
Q$ of the space--like system. This condition requires $
T^2({\omega}/{Q^2})\gtrsim Q$ or $Q\lesssim Q_s(\omega,T)$, where $Q_s
\sim ({\omega} T^2)^{1/3}$ is the plasma {\em saturation momentum}.

To summarize, a virtual photon with relatively high energy, or low
virtuality $Q\lesssim Q_s(\omega,T)$, disappears into the plasma because
this photon can branch into partons under the action of the plasma force.
Then the daughter partons can branch again and again, until the
virtuality of the descendents along this partonic cascade is reduced to a
small value of order $T$. When this happens, the fluctuation has {\em
thermalized} : the partons become a part of the thermal bath.

A similar branching picture holds for the quanta emitted by a heavy quark
which propagates with constant (average) velocity $\upsilon$ through the
plasma \cite{String}. The virtual quanta with very high virtuality $Q\gg
Q_s$ do not interact with the plasma and thus are reabsorbed by the
quark: they are a part of its wavefunction. The low virtuality quanta
with $Q\lesssim Q_s$ can decay under the action of the plasma force and
thus initiate partonic cascades which eventually thermalize (cf.
Fig.~\ref{fig:quench} right). As before, $Q_s \sim ({\omega} T^2)^{1/3}$
depends upon the energy $\omega$ of the emitted quanta, which in turn is
constrained by $\omega/Q\lesssim \gamma$, with
$\gamma=1/\sqrt{1-\upsilon^2}$ (since the quanta cannot propagate faster
than the heavy quark). The dissipation is controlled by the most
energetic among the emitted quanta. By the previous arguments, these are
the quanta with virtuality $Q\sim Q_s$ and with energy $\omega\sim
Q_s\gamma$. The last condition together with $Q_s \sim ({\omega}
T^2)^{1/3}$ imply $(Q_s)_{\rm max}= \sqrt{\gamma}\, T$. By also recalling
that it takes a time $\Delta t\sim \omega/Q^2$ to emit a quanta with
energy $\omega$ and virtuality $Q$, we conclude that
  the rate for energy loss can be estimated as follows:
  \begin{eqnarray}\label{dEdt}
 -\,\frac{{\rm d} E}{{\rm d} t}\,\simeq\,\sqrt{\lambda}\,
 \frac{ \omega}{(\omega/Q_s^2)}
 \,\simeq\,\sqrt{\lambda}\,Q_s^2 \,\sim\,\sqrt{\lambda}\,\gamma\,T^2
 \,.\end{eqnarray}
The factor $\sqrt{\lambda}$ expresses the fact that, at strong coupling,
the heavy quark radiates a large number of quanta, $\sim
{\sqrt{\lambda}}$, in the time interval $\Delta t$.

Note that this mechanism for energy loss is different from the dominant
respective mechanism as weak coupling. In both cases we can speak of
medium--induced radiation. But the dynamics allowing for this radiation
is different at weak coupling, where it involves medium rescattering (cf.
Fig.~\ref{fig:quench} left), and respectively strong coupling, where it
proceeds via medium--induced parton branching (cf. Fig.~\ref{fig:quench}
right). In both cases, this mechanism also provides a transverse momentum
broadening. At weak coupling, this has been estimated in
Eq.~\eqref{qhat}. At strong coupling, this is generated by the random
recoils associated with successive emissions, which are independent from
each other. Each such an emission gives a maximal recoil $k_\perp\sim
Q_s$ and their effects add in quadrature. This yields
 \begin{eqnarray}\label{dpTdt}
 \frac{{\rm d} \langle k_\perp^2\rangle}{{\rm d} t}\,\sim\,
\frac{\sqrt{\lambda}\,Q_s^2}{(\omega/Q_s^2)} \,\sim\,
\sqrt{\lambda}\,\frac{Q_s^4}{\gamma Q_s}\,\sim\,
 \sqrt{\lambda}\,\sqrt{\gamma}\,T^3\,.\end{eqnarray}
Eqs.~\eqref{dEdt} and \eqref{dpTdt} are consistent with the respective
AdS/CFT results, as obtained from the trailing string
\cite{Energyloss,Broadening}. In these calculations, the saturation
momentum $Q_s= \sqrt{\gamma}\, T$ appears as an induced horizon on the
string world--sheet, at $\chi_h= 1/Q_s$. In that context, the string
fluctuations are generated as `thermal noise' at this induced horizon,
with effective temperature $T=1/z_h=Q_s$ \cite{String}. This is similar
to the Unruh effect in general relativity.

So far, we have assumed the medium to be `infinite' (meaning much larger
than the formation time $\Delta t\sim \omega/Q^2$ of a virtual quanta),
but this is generally not the case in realistic heavy ion collisions.
There a `hard probe' like the heavy quark is produced as a {\em bare}
parton via a hard scattering inside the medium. The time it takes that
bare parton to dress itself with virtual quanta is generally larger (at
least for the softer quanta) than the size $L$ of the medium. Hence these
quanta feel the effect of the plasma force $F\sim T^2$ over a time
interval of order $L$, and not $\omega/Q^2$. Accordingly, the
`saturation' condition (mechanical work $\gtrsim\, Q$) becomes $Q\lesssim
T^2 L\equiv Q_s$. With this new expression for $Q_s$, the previous
estimates for the rate for energy loss and momentum broadening become
\cite{String}
 \begin{eqnarray}\label{dEdtL}
 -\,\frac{{\rm d} E}{{\rm d} t}
 \,\sim \sqrt{\lambda}\,Q_s^2 \,\sim
 \sqrt{\lambda}\,L^2 T^4\,,\qquad
  \frac{{\rm d} \langle k_\perp^2\rangle}{{\rm d} t}\,\sim
  \sqrt{\lambda}\ \frac{Q_s^2}{L}
 \, \sim \sqrt{\lambda}\,L\,T^4\,.
\end{eqnarray}
These equations imply $\Delta E\propto L^3 T^4$ and $\langle
k_\perp^2\rangle \propto L^2 T^4$, which show an enhanced medium
dependence (by a factor $LT\gg 1$) as compared to the corresponding
estimates at weak coupling \cite{Wiedemann:2009sh}.


\end{document}